\documentclass[useAMS,usenatbib]{mn2e}
\usepackage{amsmath, amssymb}
\usepackage{graphics}
\usepackage{epsfig}

\begin{document}

\title[Tidal Disruption of Stars as an EM Signature of Black Hole Recoil]{Prompt Tidal Disruption of Stars as an Electromagnetic Signature of Supermassive Black Hole Coalescence}
\author[N. Stone and A. Loeb]{Nicholas Stone$^1$ and Abraham Loeb$^1$\\
$^1$Harvard-Smithsonian Center for Astrophysics, 60 Garden Street, Cambridge, MA 02138, USA}

\pagerange{\pageref{firstpage}--\pageref{lastpage}} \pubyear{2010}

\maketitle

\label{firstpage}

\begin{abstract}
A precise electromagnetic measurement of the sky coordinates and
redshift of a coalescing black hole binary holds the key for using its
gravitational wave (GW) signal to constrain cosmological parameters
and to test general relativity.  Here we show that the merger of $\sim
10^{6-7}M_\odot$ black holes is generically followed by electromagnetic flares from tidally disrupted stars.
The sudden recoil imparted to the merged black hole by GW emission
promptly fills its loss cone and results in a tidal disruption rate of
stars as high as $\sim 0.1~{\rm yr}^{-1}$.  The prompt
disruption of a star within a single galaxy over a short period
provides a unique electromagnetic flag of a recent black hole
coalescence event, and sequential disruptions could be used on their own to calibrate the
expected rate of GW sources for pulsar timing arrays or the proposed
{\it Laser Interferometer Space Antenna}.
\end{abstract}

\label{lastpage}


\section{Introduction}
Recently, general relativistic simulations demonstrated that the
coalescence of a black hole binary is accompanied by the anisotropic
emission of gravitational radiation, causing a typical recoil of
hundreds of ${\rm km~s^{-1}}$ for the black hole remnant
\citep{Pre, Bak, Cam}. Binaries of supermassive black holes (SMBH) are a generic
consequence of galaxy mergers \citep{CD, Esc, May, Cal}, and the resultant
gravitational waves (GWs) are potentially detectable with the proposed
{\it Laser Interferometer Space Antenna} (LISA)
\footnote{http://lisa.nasa.gov/} or existing Pulsar Timing Arrays
(PTAs) such as NANOGrav \footnote{http://nanograv.org/}.

LISA would be most sensitive to binary mergers with a total mass
$M_{\rm BH}\sim 10^{6-7}M_{\odot}$ \citep{McW}, whereas the PTA
sensitivity peaks at $\sim 10^8 M_{\odot}$ \citep{SVV}.  PTAs have a
significantly poorer localization ability, with a typical uncertainty\citep{SV}
of $\sim 40$ compared to $\lesssim 1$ square degrees for LISA.  These
positional errors limit the precise determination of the luminosity
distance to merging binaries.  An electromagnetic (EM) counterpart
would greatly reduce the positional error to sub-arcsecond scales, and
also determine the redshift of the source, which would enable its use
as a ``standard siren'' (independent of the cosmic distance ladder)
for precision measurements of the dark energy equation of state
\citep{HH, Aru, Sch, Bod}.

For these reasons, prompt EM signals are of primary importance for
realizing the full potential of GW cosmology. The proposed prompt EM
signals have so far all assumed the uncertain presence of a
circumbinary accretion disk prior to coalescence.  Dissipation of GW
energy in the disk might result in a weak EM transient shortly after
the merger \citep{KL}, re-equilibration of the inner edge of the
disk could create an X-ray brightening on a timescale of $10$--$10^3
{\rm ~yr}$ \citep{MP}, and shocks produced by the GW-induced
recoil might generate EM reverberations after the recoil \citep{LFH}
which may take $\sim 10^4$ years to dissipate as enhanced infrared
luminosity \citep{SK}. It is not obvious whether these EM
signals could be distinguished from the much more abundant sources of
temporal variability in single SMBH quasars.  Moreover, the luminosity
of any circumbinary disk is expected to be significantly reduced by
the cavity associated with the decoupling of the binary from the inner
edge of the disk in the final stage of
inspiral \citep{MP, SK}.  The disk is not expected to
refill the cavity and return to its full luminosity for a time $\sim
7(1+z)(M_{\rm BH}/10^6M_{\odot})^{1.32} \rm ~yr$ after
coalescence \citep{MP}.  On longer timescales, the portion of
the accretion disk that remains bound to the recoiled SMBH is expected
to be detectable as a kinematically and eventually a spatially offset
quasar \citep{Loe, BSS, SB, KZL, Com, Shi}, although its lifetime is limited by the supply of
gas that can remain gravitationally bound to it \citep{BL}.

Here we show that the tidal disruption of stars provides a prompt EM
flag that does not depend on the prior existence of a gaseous disk in
the vicinity of the merging binary, and can result from mergers of
gas-poor galaxies. Recent observations have demonstrated that tidal
disruption events (TDEs) have a generic lightcurve and emission
spectrum \citep{Gez1, Gez2, Gez3, Don, Esq} that are distinguishable from normal quasar
variability.  Moreover, we find that SMBH recoil results in a
sequence of TDEs over a timescale of decades and potentially years, with a rate
that is $\sim 4$ orders of magnitude higher than the typical TDE
rate in normal galaxies \citep{Don}.  The existence of TDEs
accompanying SMBH mergers has been studied in the past only for long
time delays ($\sim 10^6$--$10^9$yr) before \citep{Iva, Che} or after
\citep{KM} the binary coalescence event.  While previous studies
have focused on mechanisms to slowly feed stars into an empty loss
cone, here we show that {\it GW recoil will instantaneously shift
the loss cone to a non-empty region of phase space}.

\section{Physics of the Loss Cone}
A star will be tidally disrupted by a SMBH of mass $M_{\rm BH}$ if it
passes within the tidal distance,
\begin{equation}
r_{\rm t}=r_\star (\eta^2M_{\rm BH}/m_\star)^{1/3},
\end{equation}
where $m_\star$ and $r_\star$ are the mass and radius of the star and
$\eta$ is a dimensionless constant of order unity \citep{Die}.  In
our discussion we adopt $\eta=1$ and assume an approximate main sequence scaling law of $r_* \propto m_*^{0.8}$, which implies $r_{\rm t} \propto m_*^{0.467}$\citep{MT}.
Tidal disruption does not occur if $r_{\rm t}$ is smaller than the
event horizon radius $r_h$, in which case the star is swallowed intact
by the black hole.  For non-spinning black holes and solar mass
stars, TDEs are therefore possible for $M_{\rm BH}\lesssim
10^8M_{\odot}$.  A significant black hole spin can allow for 
(angle-dependent) TDEs by SMBHs with $M_{\rm BH}\lesssim 7\times
10^8M_{\odot}$ \citep{Bel}.  During a TDE, half of the star's
mass becomes unbound, while the rest flows on Keplerian trajectories,
until the associated gas streams return to pericenter and
collisionally shock each other \citep{Ree, EK}.  These gas streams return
at a characteristic mass infall rate $\dot{M} \propto t^{-5/3}$ (roughly speaking; see \cite{LKP} for a more thorough treatment) and
form an accretion disk whose blackbody emission peaks in the UV or
soft X-ray with luminosities comparable to a supernova \citep{LU, SQ}.
Other sources of emission include line radiation from the unbound
debris \citep{KR}, and a possible brief period of super-Eddington
mass fallback. These features are useful for differentiating TDE
flares from supernovae or quasar variability, and some have already been applied to candidate events\citep{Kom}.

In a spherical galaxy with a stationary SMBH, a star is tidally
disrupted if its orbital angular momentum per unit mass falls
below a critical value,
\begin{equation}
J^2=|{\bf x} \times {\bf v}|^2<J^2_{\rm crit}\approx 2GM_{\rm BH}r_{\rm
t},
\end{equation}
where we have approximated the relevant orbit as nearly radial.
Such orbits define the so-called ``loss cone''.  The rate of TDEs for
a stationary SMBH is set by relaxation processes. Inward of a certain
galacto-centric radius the loss cone is empty, but outside of it there
is a ``pinhole'' regime where the rate of scatter in and out of the
loss cone is greater than the orbital frequency \citep{MT}.

Over the long orbital timescale of stars, the impulsive GW-induced
recoil of the SMBH remnant from a binary coalescence event yields a
nearly instantaneous change in the black hole velocity relative to the
stars \citep{OL}. Viewed from the rest frame of the black hole, there is a sudden
shift in the velocity of all stars, yielding a new loss cone defined
by
\begin{equation}
J^2=|{\bf x} \times ({\bf v}- {\bf v}_{\rm k})|^2<J^2_{\rm crit}\approx
2GM_{\rm BH}r_{\rm t},
\label{Eq2}
\end{equation}
where ${\bf v}_{\rm k}$ is the SMBH kick velocity.

We parametrize the density of stars around a SMBH binary in the last
stages of its inspiral, using a power-law profile,
\begin{equation}
\rho=\rho_0(r/r_0)^{-\gamma}.
\end{equation}
This density profile corresponds to an isotropic pre-kick distribution function 
of stars,
\begin{equation}
f(r, v)=C(2GM_{\rm BH}/r-v^2)^{\gamma-3/2},
\end{equation}
where the normalizing constant is given by
\begin{equation}
C=\frac{(3-\gamma)(\gamma-0.5)\Gamma(\gamma+1)}{2\pi^{2.5}
\Gamma(\gamma+0.5)}\frac{M_{\rm BH}}{r_{\rm inf}^3}
\left(\frac{r_{\rm inf}}{2GM_{\rm BH}}\right)^{\gamma},
\end{equation}
which in turn depends on the observationally calibrated \citep{MSK}
radius of influence of the SMBH (inside of which the mass in stars is
$2M_{\rm BH}$).  For a full sample of galaxies this is
\begin{equation}
r_{\rm inf}=24 (M_{\rm BH}/10^8M_{\odot})^{0.51}~{\rm pc},
\label{RInfGeneral}
\end{equation}
while for core galaxies alone it is
\begin{equation}
r_{\rm inf}=35 (M_{\rm BH}/10^8M_{\odot})^{0.56}~{\rm pc}.
\label{RInfCore}
\end{equation}
Stars will be bound to the black hole system of total mass $M_{\rm
BH}$ before the binary coalescence if
\begin{equation}
{\bf v}^2<2GM_{\rm BH}/r, 
\end{equation}
and after coalescence if
\begin{equation}
({\bf v}-{\bf v}_k)^2<2GM_{\rm BH}/r.
\end{equation}
The intersection of these two spheres in velocity space, with each
other and with the loss cone, is the region of phase-space containing
bound stars which are tidally disrupted after the recoil.  By
performing a Monte Carlo integration of the appropriate distribution
function over this region we calculate the number of post-recoil TDEs.
Through a separate integral we find that the unbound stars provide
$\lesssim 10\%$ of the total number of TDEs and can be neglected.
Another small correction to the total rate involves the net SMBH mass
loss by GW emission, which is typically $\lesssim 5\%$ of the pre-merger
mass \citep{Cam}.  We have included the associated small reduction (by
$\sim 10\%$) in the number of tidally disrupted stars.  To evaluate
the observability of the recoil-induced TDEs, we define the quantity
$N_<(t)$ as the number of stars in the post-recoil loss cone which are
tidally disrupted in $t$ years.  Since stars which fall into the new
loss cone are near their apocenter in the rest frame of the kicked
black hole, $N_<(100)$ is the number of stars with orbital periods below
200 years.

The result of the Monte Carlo integral is sensitive to the innermost
region of the distribution function, whose details are determined by
the pre-merger dynamical environment.  In particular, the SMBH binary
will excavate a larger and more complex loss region than is given by
Eq. (\ref{Eq2}).  Below, we consider the innermost regions of phase
space in gas-poor and gas-rich mergers, separately.

\section{Dry Mergers}

\begin{figure}[!t]
\includegraphics[width=84mm]{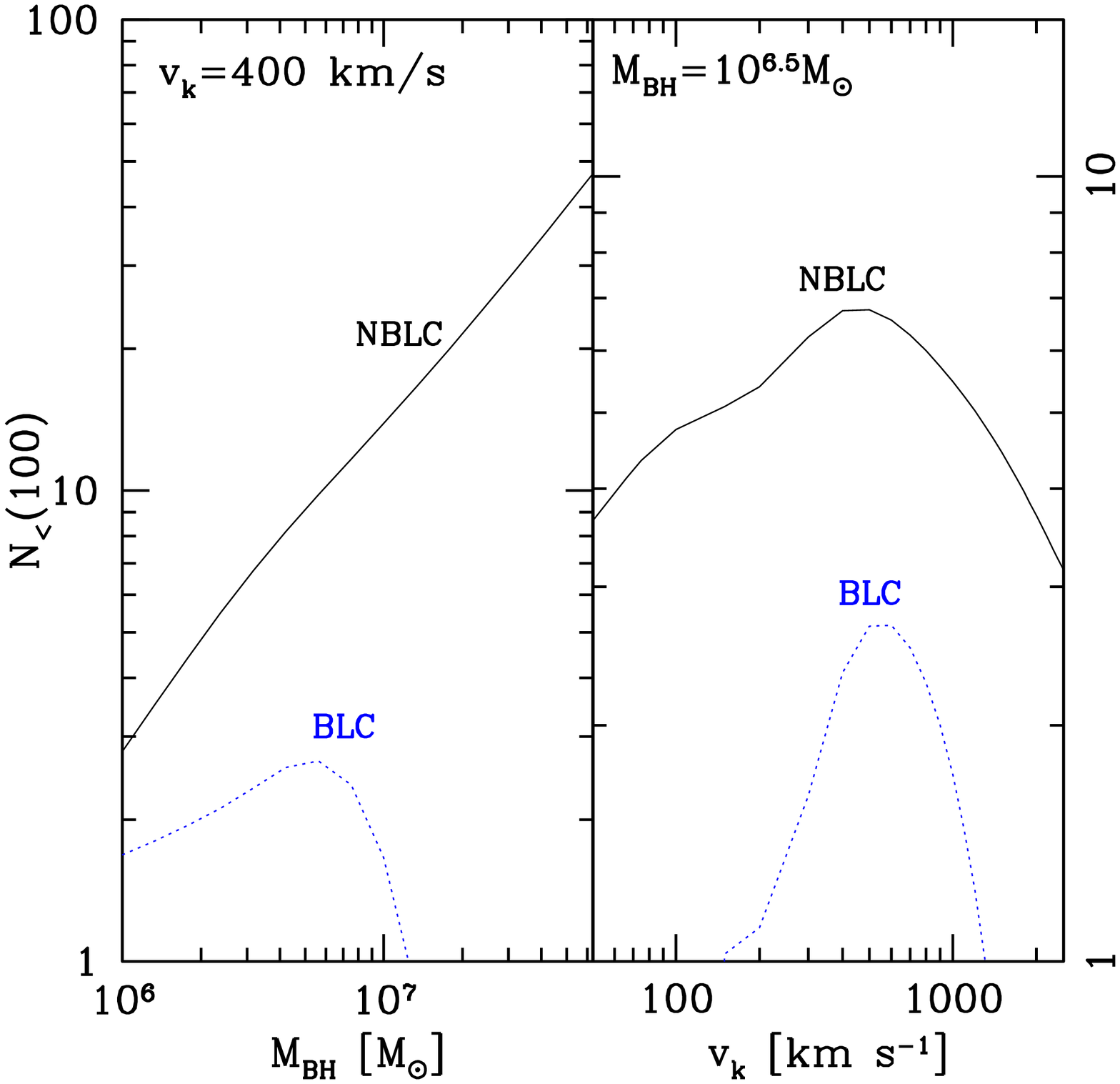}
\caption{Expected number of stars disrupted in less than 100 years,
$N_<(100)$, in the dry merger, joint core-cusp case.  {\it Left
panel:} Black hole mass dependence for kick velocity $v_{\rm
k}=400~{\rm km~s^{-1}}$.  The dotted blue line (labeled `BLC')
represents the physically realistic scenario including the binary loss
cone discussed in the text, whereas the black line (labeled `NBLC')
accounts only for the loss cone of a single black hole.  {\it Right
panel:} Kick velocity dependence for a black hole mass $M_{\rm
BH}=10^{6.5}M_{\odot}$, using the same line types.}
\label{DryBoth}
\end{figure}

In the absence of gas, $\gamma=1.75$ is the dynamically relaxed
(so-called ``Bahcall-Wolf'') equilibrium state \citep{BW} of a
stellar cluster around a SMBH.  However, core galaxies are believed to
be the end product of a binary inspiral, as the binary sheds angular
momentum by ejecting stars.  Numerical simulations of this process
show that a binary hardening its orbit through scattering of stars
will scour a core \citep{Mer}, but at some point depletion of the
remaining stars in the binary loss cone (including in this context
stars whose pericenters fall within twice the binary semimajor axis)
will lead to a stalling of the binary and the so-called ``final parsec
problem'' \citep{MM}.  Without gas, the binary can only merge via a
re-population of its loss cone.  Significant triaxiality of the galaxy
potential \citep{MPo} tends to re-populate the binary loss-cone but
preserve a core with $\gamma \approx 1$.  Alternatively, collisional
processes generate a central cusp of stars, though this method of
binary hardening is only effective for binary masses $\lesssim 10^7
M_{\odot}$ \citep{MSK}, and it may be modified in the presence of massive perturbers\citep{PA}, such as infalling molecular clouds in gas-rich mergers.  These gas-free scenarios lead us to consider
both core galaxies where $\gamma =1$, and galaxies with a joint
core-cusp density profile where an inner $\gamma =1.75$ profile meets
an outer $\gamma =1$ profile at a radius of $0.2r_{\rm inf}$ \citep{MSK}.  We use Equation(\ref{RInfCore}) as the radius of influence in these models.

It is necessary to exclude the stars located in the pre-coalescence
loss cone.  The size of this region of phase space is somewhat
uncertain.  For collisional re-population of the loss cone, numerical
simulations \citep{MMS} indicate the SMBH binary will decouple from a
relaxed distribution of stars at a semimajor axis of $a_{\rm eq}\sim
10^{-3}r_{\rm inf}$, and coalesce before the stars can relax into the
gap left behind.  To model this cavity in energy space we remove all
stars with semimajor axes less than this radius.  However, because
relaxation in angular momentum is faster than in energy, the resultant
gap in angular momentum space will be partially refilled prior to the
merger.  The timescale for filling up a gap in angular momentum space
is given by \citep{MW},
\begin{equation}
T_{\rm gap}=\frac{a}{r_{\rm inf}}T_{\rm r},
\end{equation}
where $a$ is the semimajor axis of the SMBH binary, taken to be the
pericenter at which stars are ejected \citep{MW}, and the system
relaxation time at the radius of influence is \citep{MSK},
\begin{equation}
T_{\rm r}\approx 8.0\times 10^9 {\rm yr} \left( \frac{M_{\rm BH}}{10^6
M_{\odot}}\right)^{1.54} .
\end{equation}
Thus, a second cavity in the distribution function is created by
removing all stars with pre-kick pericenters lower than the
binary separation $a$ at which $T_{\rm gap}$ equals the gravitational
wave timescale,
\begin{equation}
T_{\rm GW}=\frac{5c^5a^4}{256G^3M_{\rm BH}^2\mu},
\end{equation}
with $\mu=M_1M_2/M_{\rm BH}$ being the reduced mass of the binary and
$M_{\rm BH}=(M_1+M_2)$.  For simplicity, we adopt a flat $J$ dependence for $f(E,J)$ with the cuts mentioned above.  In the classical loss cone calculation, the steady state solution of the orbit-averaged Fokker-Planck equation  yields a distribution function that varies logarithmically with $J$ at fixed $E$ \citep{CK}.  However, this solution may not apply to the innermost stars, for which the orbit averaged assumption may break down and strong star-star scatterings (which the Fokker-Planck approach does not account for) could be important. Furthermore, the loss cone of a binary SMBH is not the pure sink assumed for a single BH, since stars may remain bound to the binary on low-$J$ orbits.  The use of a logarithmic instead of a flat distribution would have reduced $N_<(t)$ by a factor of $\sim 2-4$. 

We use these cuts in stellar energy and
angular momentum as modifications to the joint core-cusp profile. We
generalize the results to a range of stellar masses using a Salpeter
initial mass function (IMF) with a differential number of stars,
${dN_\star}/{dm_\star} \propto m_\star^{-2.35}$ in the mass range
$0.1M_{\odot}<m_\star<100M_{\odot}$.  The number of disruptions is dominated by
low mass stars despite their smaller $r_{\rm t}$; switching to a
top-heavy IMF, as is sometimes discussed in the context of galactic
nuclei \citep{Bar}, would reduce $N_<(100)$ by a factor of a few.

\section{Wet Mergers}
In gas-rich mergers, the pre-kick profile is likely to be different.
On the one hand, rapid loss of angular momentum by the binary to
dynamical friction on the gas can produce a core by denying stars the
time needed to relax into a central cusp as described above
\citep{MSK}; but on the other hand, {\it in situ} star formation could
rebuild a nuclear cusp while the binary orbit hardens.  The
possibility of star formation, and subsequent migration, in disks
motivates us to consider values of $\gamma=1.5, 1.75, 2$.
Alternatively, stars formed elsewhere could be ``ground down'' into
orbits inside the disk \citep{SCR}, and then behave in a similar
fashion.  Although the details of star formation in disks fragmenting
due to gravitational instability are quite complex \citep{SBe, AAC}, we
provide an approximate description of their potential to contribute to
the post-kick loss cone here.  If the Roche radius of the star in the
disk exceeds the disk scale height, and tidal coupling of the star to
the disk is at least as effective as viscosity at transporting angular
momentum, the star will open a gap in the disk \citep{SCR} and migrate
inward on a viscous timescale, to the point where those conditions are
no longer met, or the disk's inner edge \citep{GT}, whichever is
larger.  Here we use Equation (\ref{RInfGeneral}) for the radius of influence, since we are considering galactic nuclei in the process of rebuilding their cusps.  

\begin{figure}[!t]
\includegraphics[width=84mm]{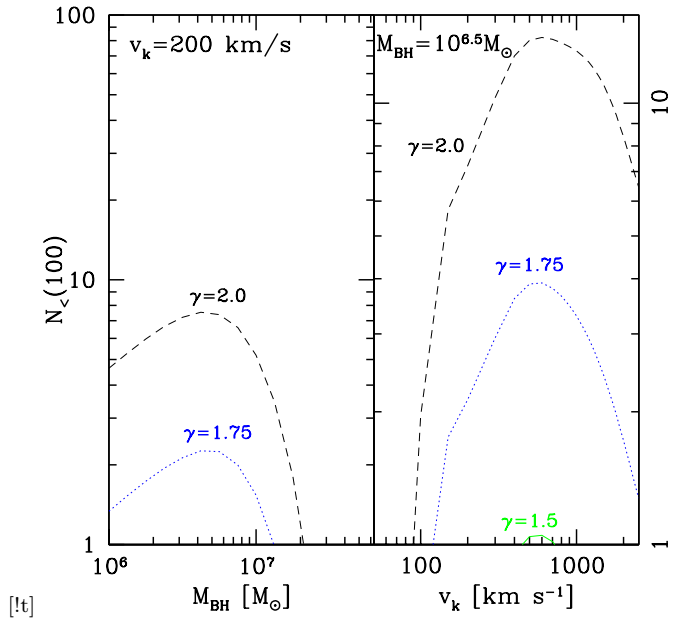}
\caption{Expected number of stars disrupted in less than 100 years, $N_<(100)$, in the wet merger scenario.  {\it Left panel:} Mass dependence with $v_{\rm k}=200~{\rm km~s^{-1}}$.  The solid green line represents a $\gamma=1.5$ cusp, the dotted blue $\gamma=1.75$, and the dashed black $\gamma=2.0$.  {\it Right panel:} Velocity dependence with $M_{\rm BH}=10^{6.5}M_{\odot}$, and the same lines as in the left panel.}
\label{WetBoth}
\end{figure}

We follow a similar procedure as with dry mergers, to approximate the
size of the pre-merger loss cone.  Assuming a thin disk, the radial
size of the central cavity is determined by setting the viscous
timescale at the radius of marginal self-gravity \citep{GT},
\begin{equation}
T_{\rm vis} = 4.2 \times 10^5 {\rm yr} ~\alpha_{0.3}^{-1/3}\kappa^{-1/2}\mu^{1/3}\left(\frac{\epsilon_{0.1}}{l_E}\right)^{1/6}M_8^{1/2},
\end{equation}
equal to $T_{\rm GW}$.  Here, $\kappa$ is the opacity in units of
electron-scattering opacity, $\mu$ is mean gas particle mass in units
of the proton mass, $\alpha_{0.3}$ is the standard (Shakura-Sunyaev)
viscosity parameter scaled to 0.3, $\epsilon_{0.1}$ is the radiative
efficiency scaled to $10\%$, and $l_E$ is the total radiated
luminosity in units of the Eddington limit for the black hole mass
$M_{\rm BH}=M_8\times 10^8M_\odot$.  Noting the weak power law
dependences in $T_{\rm vis}$, we set all parameters except $M_{\rm
BH}$ to their fiducial values \citep{GT}.  We remove any stars with
pericenters interior to the radius at which the SMBH binary decouples
from the accretion flow, and assume that the remaining stars obtain a
nearly spherical angular distribution (similarly to the innermost
S-stars in the Milky-Way nucleus \citep{Ghe}) before the SMBH recoil.
Because the details of star formation within the disk are highly
uncertain and the TDE rate is dominated by low-mass stars, we assume
for simplicity $m_\star\sim 1M_\odot$ in the wet merger case.  We
emphasize that in this case, accretion-induced alignment of SMBH spins
prior to merger is expected to strongly suppress kicks over $~200 {\rm
km~s^{-1}}$ \citep{Bog, Dot}.

\section{Other Considerations}
Other processes could also partially refill the binary loss cone.  In
analogy to the problem of resonant capture during planetary migration
\citep{YT}, mean-motion resonances could be capable of pulling stars
inward during the final stages of the SMBH merger, dramatically
increasing the number of post-kick disruptions.  The special case of
the 1:1 Lagrange point resonance has recently been investigated
\citep{Scn} and found capable of migrating stars to within tens
of Schwarzschild radii from the system barycenter \citep{SM}.  Higher integer ratio mean-motion resonances have been seen to affect stars (see \cite{Che}, Figures 5, 6, 7) as a binary SMBH hardens, although their ability to drive resonant migration is less clear.
A detailed study of resonant migration in SMBH binaries is
beyond the scope of this paper, but this effect has the potential to
dramatically expand the short-period population of the post-kick loss cone.

For a source at a redshift $z$, cosmological time dilation will
stretch the duration of each TDE flare, delay the onset of the first
post-kick flare, and reduce the observed TDE rate all by the same
factor of $(1+z)$.  However, observations suggest that mean density of
stars in high-redshift galaxies scales as $(1+z)^3$ \citep{Oes}.  This
could lead to a net enhancement in the observed TDE rate $\propto
(1+z)^2$ per galaxy, if the central regions of galaxies are
self-similar.  We set $z=0$ to ignore these possible cosmological
effects in our calculated TDE rates.

\section{Results}

\begin{figure}[!t]
\includegraphics[width=84mm]{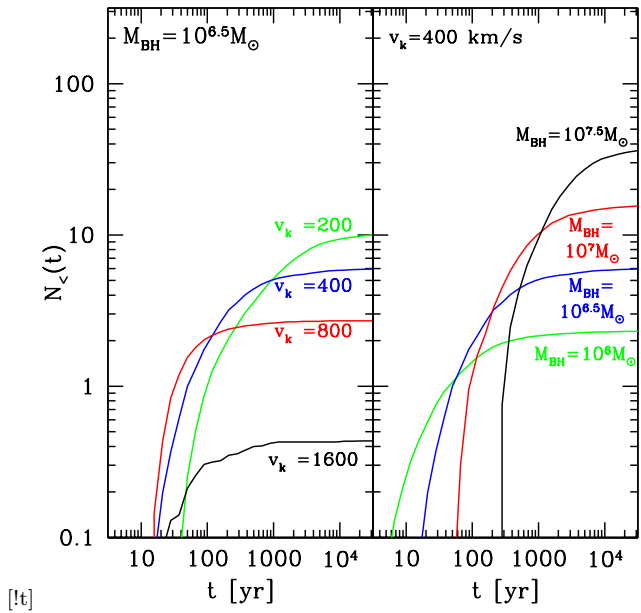}
\caption{{\it Left panel:} Expected number of stars disrupted in less than $t$ years, $N_<(t)$,
for the dry merger model and $M_{\rm BH}=10^{6.5}M_{\odot}$.  The green line is for $v_{\rm k}=200~{\rm km~s^{-1}}$, the blue line is for $v_{\rm k}=400~{\rm km~s^{-1}}$, the red line is for $v_{\rm k}=800~{\rm km~s^{-1}}$, and the black line is for $v_{\rm k}=1600~{\rm km~s^{-1}}$.  {\it Right panel:} $N_<(t)$ for $v_{\rm k}=400 \rm ~km~s^{-1}$ and varying masses.  The green line is for $M_{\rm BH}=10^6M_{\odot}$, the blue for $M_{\rm BH}=10^{6.5}M_{\odot}$, the red for $M_{\rm BH}=10^7M_{\odot}$, and the black for $M_{\rm BH}=10^{7.5}M_{\odot}$.}
\label{CDFBoth}
\end{figure}

For simplicity, our calculations assume binaries of equal mass black
holes.  An unequal mass would increase $T_{\rm GW}$ and increase
$N_<(t)$ by allowing more time for refilling the binary loss cone, but
it would also decrease the likely values of $v_k$.  The latter change
dominates only for mass ratios smaller than $\sim 0.1$, so our results should
be regarded as conservative for major mergers.  Figure \ref{DryBoth} presents the
velocity and mass dependences of our most realistic model (the joint
core-cusp profile, for SMBH binaries that harden in dry mergers by
scattering of stars), as well as a less realistic core-cusp model
included for illustrative purposes.  The first model (labeled BLC)
removes stars from the binary loss cone and results in an interesting
number of TDEs for $M_{\rm BH}\lesssim 10^7M_{\odot}$ and $200 {\rm
~km~s^{-1}} \lesssim v_{\rm k} \lesssim 1000 {\rm ~km~s^{-1}}$.  At
low velocities, overlap with the pre-merger loss cone sharply
suppresses $N_<(100)$, while at high velocities, the reduced size of
the bound stellar population also shrinks $N_<(100)$.  At higher
masses, all short-period stars are scoured by the pre-merger binary
loss cone.  The second case in Figure \ref{DryBoth} (labeled NBLC)
replaces the loss cone of a binary with that of a single black hole.
Interestingly, the dramatic increase in $N_<(100)$ here results from
the addition of relatively few stars ($\sim 100$ for $M_{\rm BH}=10^6M_{\odot}$),
indicating that resonant migration of a small population of stars could
significantly boost $N_{<100}$.  Finally, the $\gamma =1$
(pure core) case does not have enough centrally located stars to
produce post-kick tidal disruptions on $<100~ \rm yr$ timescales.

The results for wet mergers are illustrated in Figure
\ref{WetBoth}. Although these rely on significantly more uncertain
assumptions than the dry merger model, they show that star formation
can produce interesting values of $N_{<100}$ in the $\gamma=1.75$ and
$\gamma=2$ cases.

Figure \ref{CDFBoth} shows how the delay until the first post-kick
disruptions changes with black hole mass and kick velocity. 
Here we consider the joint core-cusp model (with removal
of pre-merger loss cone by the binary), and find that the first
disruption is expected to occur between two and five decades after
SMBH coalescence for black holes with masses between $10^6 M_{\odot}$
and $10^7M_{\odot}$, and kick velocities between $400$ and $800~ \rm
km~s^{-1}$.  Fortunately, this region of parameter space falls within
both the black hole mass range LISA is likely to observe, and the
range of physically plausible recoil velocities for dry mergers.

\section{Summary} We find that merging galaxies 
with black hole masses $M_{\rm BH}\lesssim 10^7M_\odot$ are likely to
produce a tidal disruption flare a few decades after coalescence in
the case of a gas-poor merger.  Multiple flares could
possibly be seen on a timescale of years if resonant migration is
effective. For gas-poor mergers, the peak rate is therefore at least
$\sim 10^4$ times higher than the typical TDE rate in galaxies
\citep{Don}. The total number of TDEs is maximized, and delay until
the first TDE minimized, if the kick velocity is in the range of
$200$--$1000 \rm km~s^{-1}$.

Our minimal predictions concerning dry mergers could be dramatically enhanced if  resonant migration increases the number and frequency of post-kick tidal disruptions.  The
case of wet mergers is substantially more complicated, and because
star formation and disk migration can also significantly increase the
population of the post-kick loss cone, a much more detailed study of
star-disk interactions around binary SMBHs is needed to make robust
predictions in the gas-rich case.

Moderate to high kick velocities in dry mergers will provide a robust
EM counterpart to the GW signature of black hole coalescence within
the LISA band, enabling accurate identification of the host galaxy and
a precise measurement of cosmological parameters \citep{HH} within a
few decades of the initial GW signal.  With the advent of massive
transient surveys, such as PTF
\footnote{http://www.astro.caltech.edu/ptf/}, Pan-STARRS
\footnote{http://pan-starrs.ifa.hawaii.edu/public/}, and LSST
\footnote{http://www.lsst.org/lsst}, it is possible that sequential tidal disruption
flares could flag black hole recoil events without a GW signal,
providing an independent test of the strong field regime of general
relativity and a calibration of the expected event rate for LISA and
PTAs.  

\bigskip
\section*{Acknowledgments.}
We thank Matt Holman, Bence Kocsis, Ryan O'Leary, Hagai Perets, and Alberto Sesana for helpful comments on the manuscript. This work was
supported in part by NSF grant AST-0907890 and NASA grants NNX08AL43G
and NNA09DB30A.

\end{document}